# Unprecedented Enhancement of Piezoelectricity in Wurtzite Nitride Semiconductors via Thermal Annealing

Shubham Mondal[1*], Md Mehedi Hasan Tanim[1*], Garrett Baucom[2], Shaurya S. Dabas[3], Jinghan Gao[3], Venkateswarlu Gaddam[3], Jiangnan Liu[1], Aiden Ross[4], Long-Qing Chen[4], Honggyu Kim[2], Roozbeh Tabrizian[3] and Zetian Mi[1, *]

[1]Department of Electrical Engineering and Computer Science, University of Michigan, Ann Arbor, MI 48109, USA

[2]Department of Materials Science and Engineering, University of Florida, Gainesville, FL 32611, USA

[3]Electrical and Computer Engineering Department, University of Florida, Gainesville, FL 32611, USA

[4]Department of Materials Science and Engineering, The Pennsylvania State University, University Park, Pennsylvania 16802, USA

*E-mail: ztmi@umich.edu

*Abstract:* The incorporation of rare-earth elements in wurtzite nitride semiconductors, *e.g.,* scandium alloyed aluminum nitride (ScAlN), promises dramatically enhanced piezoelectric responses, critical to a broad range of acoustic, electronic, photonic, and quantum devices and applications. Experimentally, however, the measured piezoelectric responses of nitride semiconductors are far below what theory has predicted. Here, we show that the use of a simple, scalable, post-growth thermal annealing process can dramatically boost the piezoelectric response of ScAlN thin films. We achieve a remarkable 3.5-fold increase in the piezoelectric modulus, $d_{33}$ for 30% Sc content ScAlN, from 12.3 pC/N in the as-grown state to 45.5 pC/N, which is eight times larger than that of AlN. The enhancement in piezoelectricity has been unambiguously confirmed by three separate measurement techniques. Such a dramatic enhancement of $d_{33}$ has been shown to impact the effective electromechanical coupling coefficient $k_t^2$ : increasing it from 13.8% to 76.2%, which matches the highest reported values in millimeter thick lithium niobate films but is achieved in a 100 nm ScAlN with a 10,000 fold reduction in thickness, thus promising extreme frequency scaling opportunities for bulk acoustic wave resonators for beyond 5G applications. By utilizing a range of material characterization techniques, we have elucidated the underlying mechanisms for the dramatically enhanced piezoelectric responses, including improved structural quality at the macroscopic scale, more homogeneous and ordered distribution of domain structures at the mesoscopic scale, and the reduction of lattice parameter ratio (*c/a*) for the wurtzite crystal structure at the atomic scale. Overall, the findings present a simple yet highly effective pathway that can be extended to other material families to further enhance their piezo responses.

**Keywords:** Piezoelectricity, ScAlN, $d_{33}$, annealing

## I. Introduction

Piezoelectricity is the generation of internal electric field in certain materials under applied mechanical stress.[1] In crystalline materials with a lack of inversion symmetry, piezoelectricity is produced due to electrochemical interaction between the electrical and mechanical states of the constituent atoms. The piezoelectric effect is also reversible, *i.e.,* a piezoelectric material can generate mechanical strain when subject to an applied electric field.[2] Since the first demonstration of direct piezoelectricity effect by P. Curie and J. Curie in 1880, they have been widely utilized in applications such as high-voltage and high-power sources, sensors, actuators, time and frequency control, surgery, and ultrasound medical devices.[1,3] Thin film piezoelectric materials have also been explored and intensively studied to advance a broad range of emerging devices and systems, including next-generation microelectromechanical systems (MEMS)[3-6] and acoustic wave resonators and filters for future communication technologies.[7] They have already reshaped communication technologies like 5G and beyond and empowered the evolution of sensing applications such as electroacoustic and immunosensors, microphones, touch-sensitive screens, and activity monitors in smart wearables, to name just a few.[5,7-10] Additionally, as the demand for energy harvesting technologies grows, piezoelectric energy harvesting systems have gained prominence.[11]

To date, the commercial piezoelectric application space mostly relies on a few well-researched piezoelectric materials including ferroelectric ceramics such as $Pb(Zr_xTi_{1-x})O_3$ (PZT), barium titanate ($BaTiO_3$), and other perovskite titanates.[3,12] While these materials exhibit good piezoelectric properties, as measured from their longitudinal piezoelectric strain constant $d_{33,f}$ (often interchangeably used as $d_{33}$ in prior literature), they are incompatible with existing CMOS process, and their low Curie temperatures (generally below 400 °C) makes them prone to formation of oxygen vacancies, resulting in a sharp deterioration in their ferroelectric and piezoelectric properties under harsh conditions.[13,14] Additionally, PZT poses environmental hazards due to its high toxicity and volatility at high operational temperatures.[15] Recently, aluminum nitride (AlN) has emerged as a promising platform for a wide range of applications, including MEMS/nanoelectromechanical system devices as well as high frequency resonators and filters, due to its high Curie temperature (~1150 °C), minimal acoustic and dielectric losses, elevated acoustic wave velocity, non-toxicity and compatibility with back-end CMOS integration requirements.[16-18] Yet, pure AlN thin films exhibit a relatively low piezoelectric modulus ($d_{33}$ ~ 4 pC/N)[19,20], restricting its potential application in state-of-the-art piezoelectric technologies.[21] However, the introduction of one or more alloying elements in AlN can significantly enhance the piezoelectric performance.[21,22] An exhaustive list of 25+ elements have been identified that can be alloyed within AlN to enhance its piezoelectric properties.[21,23] One of the most promising approaches is to alloy AlN with Scandium (Sc) to form wurtzite $Sc_xAl_{1-x}N$, which significantly amplifies the piezoelectric constants of AlN.[19] Recent theoretical predictions reveal a maximum possible enhancement of $d_{33}$ to ~ 100 pC/N at nearly 64% Sc composition.[24] There exists discrepancies in the impact of internal strain, finite size of the super-cell and associated configurational disorder on the results, thus widely varying values have been reported for the theoretical limit of $d_{33}$ in ScAlN.[25] The significantly enhanced piezoelectric response along the *c*-axis in wurtzite ScAlN crystals, defined by the piezoelectric modulus, $d_{33}$, is primarily due to the energy landscape flattening as a result of competition between the layered hexagonal (*h*-ScAlN, attained at ~50% Sc content) and the parent wurtzite phase with increasing Sc content.[19] The addition of Sc to AlN causes distortion in the wurtzite crystal structure, resulting in a decrease in the *c*/*a* ratio and an increase in the internal *u* parameter. This shift towards a planar hexagonal structure destabilizes the crystal and enhances the piezoelectric response.

Experimentally, maintaining high-quality crystal structures becomes extremely difficult with increasing Sc content, thereby constraining the realization of exceptional piezoelectric properties predicted by theory.[30] Moreover, a high Sc content also leads to a phase transition from the wurtzite to the nonpolar rock-salt phase, which can completely eliminate piezoelectric response from the material.[19] The highest $d_{33}$ value experimentally measured in ScAlN is in the range of 25-32 pC/N, which is achieved at a Sc composition exceeding 40%, shown in **Figure 1**.[19,27,29] However, such high Sc-alloying often leads to material imperfections, limiting its practical application. Consequently, most of the ScAlN based acoustic devices reported to date are largely based on relatively lower Sc concentrations (~20 – 30%), with $d_{33}$ merely in the range of 10-15 pC/N, that is only 2-3 times larger than AlN.[26-28]

Herein, we demonstrate that a simple, scalable, post-growth thermal annealing process can dramatically enhance the piezoelectric response of ScAlN. We measure a 250% (3.5×) enhancement in $d_{33}$, from 12.3 pC/N for the as-grown $Sc_{0.3}Al_{0.7}N$ sample, to 45.5 pC/N for the sample with optimized annealing, shown in **Figure 1**. This is more than 700% (8×) higher than that of AlN and ~1.5 times higher than the highest value of $d_{33}$ reported in literature for ScAlN across all Sc compositions. This could significantly enhance acoustic device performance, as a higher $d_{33}$ value leads to more efficient mechanical-to-electrical energy conversion. This efficiency is quantified by the effective electromechanical coupling coefficient $k_t^2$, which has been demonstrated to increase substantially from 13.8% to 76.2%, approaching the highest values reported in millimeter-thick lithium niobate films but in a ScAlN film that is merely 0.01% of the thickness. This unprecedented enhancement in the piezoelectric response of ScAlN thin films unlocks its transformative potential for frequency scaling in bulk acoustic wave resonators used in centimeter and millimeter-wave microwave spectral processing and beyond 5G-wireless communication technologies.

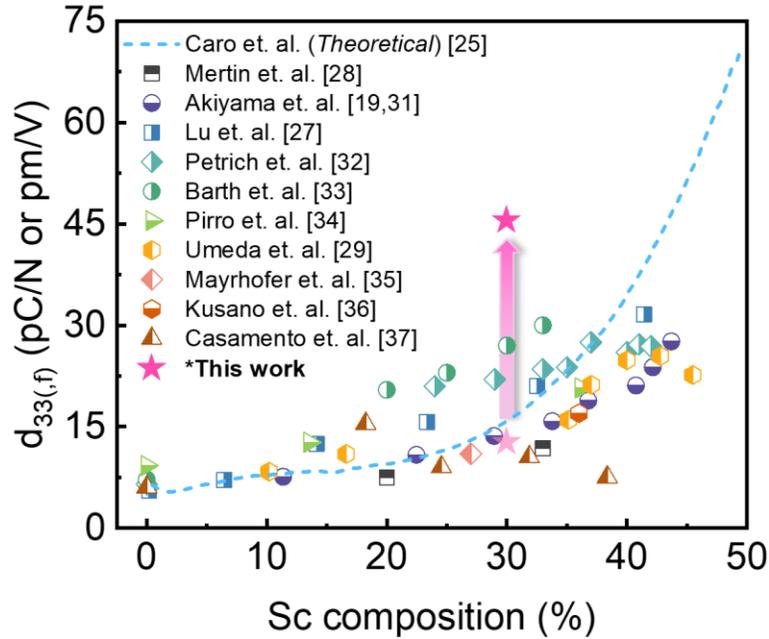

**Figure 1:** Benchmarking of piezoelectric coefficient, $d_{33}$ ($d_{33,f}$) with recent reports in literature as a function of varying Sc compositions.[19,25,27-29,31-37]

## 2. Results and Discussion

### 2.1. Thermal annealing for *d₃₃* enhancement

From the structural perspective, the piezoelectric response of wurtzite AlN depends on a plethora of factors, including but not limited to the orientation of the *c*-axis, distribution of the constituent atoms, orientation of the polar grains, M (metal) – N (Nitrogen) bond length, strain in the film and the presence of defects within the films.[38,39] Herein, we explore the potential of controlled atmosphere annealing to enhance the orientation along the *c*-axis and the overall crystallinity of the as-grown ScAlN thin films. The evolution of *d₃₃* for various annealing conditions for ScAlN are outlined in **Figure 2.** For the as-grown sample, the *d₃₃* was measured to be 12.3 pC/N from piezoelectric force microscopy (PFM) measurement, which is consistent with other reports in the literature.[32,33] Shown in **Figure 2(a)**, we observe an enhancement in $d_{33}$ with increasing annealing temperature, peaking at 45.5 pC/N at 700°C for 2 hours in vacuum, followed by a sharp decline. This marks a remarkable 3.5 times increase compared to the as-grown film. Similarly, when keeping the temperature constant at 700°C, varying the annealing duration from 1 to 8 hours reveals a gradual increase in piezo response, peaking at 45.5 pC/N after 2 hours of annealing, shown in **Figure 2(b)**. Beyond the optimal temperature, $d_{33}$ shows a sharp decline, whereas it remains reasonably stable with prolonged annealing durations. Based on the above findings, we infer that the increase in $d_{33}$ is more sensitive to annealing temperature than time. The narrow annealing window explains the nominal enhancement in $d_{33}$ achieved till date for wurtzite III-Ns despite some previous attempts.[26,40,41] Furthermore, to analyze the effect of ambient conditions on the piezo response, we varied the ambient conditions and repeated the annealing process at the optimized condition (700 °C, 2h) and the results are plotted in **Figure 2(c)**. The observed $d_{33}$ is similar for vacuum ($d_{33}$ = 45.5 pC/N) and Ar ambience ($d_{33}$ = 47 pC/N), with a reduction observed for $N_2$ ambience ($d_{33}$ = 39.8 pC/N) and a sharp reduction for $O_2$ ambience ($d_{33}$ = 34.5 pC/N).

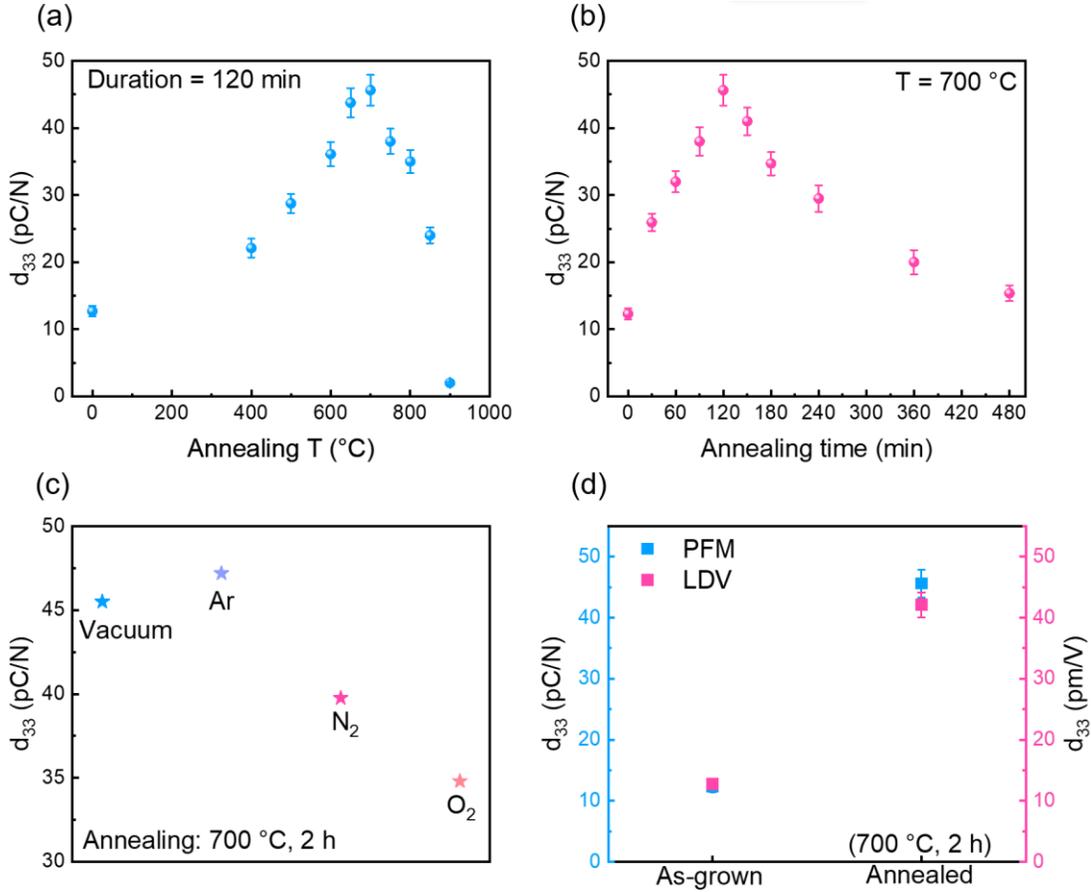

**Figure 2:** (a) Plot of $d_{33}$ versus annealing temperature showing a maximum enhancement at T = 700 °C for PVD-grown ScAlN. (b) Plot of $d_{33}$ vs annealing time showing a maximum enhancement at a duration of 2 h, showing an enhanced $d_{33\,(max)}$ = 45.5 pC/N at the optimized annealing condition from 12.3 pC/N for the as-grown sample. (c) Plot of $d_{33}$ at the optimized annealing condition in varying ambient conditions (vacuum, Ar, $N_2$ and $O_2$). (d) Plot of $d_{33}$ values, before and after annealing, showing good agreement across different measurement techniques (PFM and LDV).

Previous studies have shown that the $d_{33}$ of AlN films is highly sensitive to surface oxidation.[42] Due to the large oxygen affinity of both Sc and Al elements, and further exacerbated by the high-temperature annealing, a rapid surface oxidation process occurs when ScAlN films are annealed in an $O_2$ ambience.[43] This leads to a sharp drop in the $d_{33}$ compared to other annealing environments. Furthermore, annealing in $N_2$ and $O_2$ ambiences at a high temperature may introduce additional defects in the crystal lattice and can induce phase separations. For example, annealing AlN in excess nitrogen leads to the formation of aluminum oxynitride (Al–O–N) phase[40], which may result in a comparatively lower enhancement compared to Ar (inert) and vacuum conditions. Moreover, we carried out Laser Doppler Vibrometry (LDV) measurements on both the as-grown and optimally annealed samples. As shown in **Figure 2(d)**, the LDV results revealed a $d_{33}$ value of 12.7 pm/V for the as-grown sample and a peak $d_{33}$ value of 42.1 pm/V for the sample optimized at 700°C for 2 hours. These findings are in excellent agreement with those obtained from the PFM measurements, providing further evidence to support the dramatic enhancement achieved through annealing.

To explore correlation of electromechanical coupling with annealing, displacement and capacitance loops of the two samples were measured using a Radiant PiezoMEMS Analyzer (**Figure 3**). For displacement loop extraction, triangular pulses at 60V, 2kHz are applied to capacitors implemented in both the samples and displacement is measured by a synchronized laser vibrometer. After annealing, the largest displacement increased almost 3.5 times, from 0.5 nm to 1.7 nm, corresponding well with PFM and LDV results. The displacement loop of both samples have broad linear regions, which were used to calculate $d_{33,f}$ by fitting a line, as highlighted in **Figure 3(a)**.

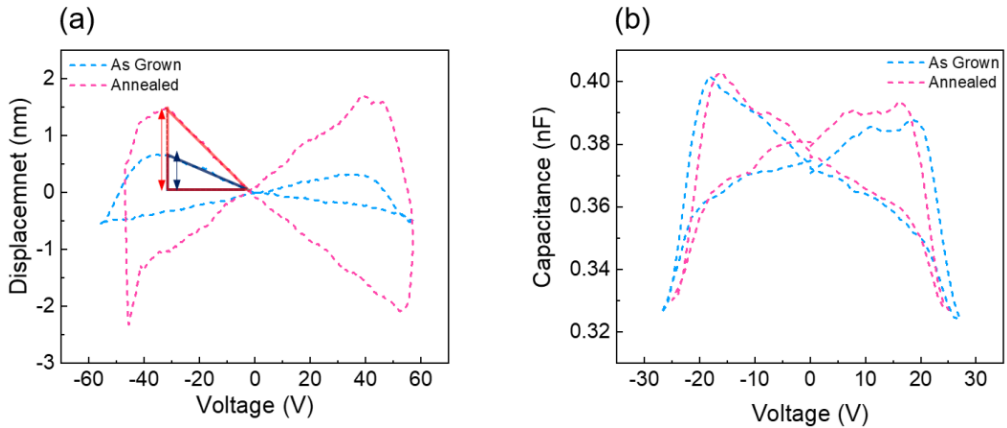

**Figure 3:** (a) Converse piezoelectric effect as measured from the displacement loop of as grown and annealed AlScN sample, from which $d_{33,f}$ could be calculated from the broad linear regions. (b) C-V loops of the as-grown and annealed film showing nearly identical response.

After annealing, an effective converse piezoelectric coefficient $d_{33,f}$ of 37.9 pm/V and -41.9 pm/V was obtained in the positive and negative branch of the loop, respectively. These results further confirm the significant enhancement of piezoelectric properties after annealing process. To enable estimation of electromechanical coupling, capacitor loops (*i.e.*, C-V) are also measured using triangular pulses at 25-27V, 1 MHz applied to 500μm×500μm capacitors implemented on the two samples (**Figure 3(b)**). A linear fit is used to extract relative permittivity ($\varepsilon_{33}$) of the samples, resulting in 16.8 and 17.1 for as-grown and annealed samples, respectively.

The extracted $d_{33,f}$ and $\varepsilon_{33}$ are used, along with *c*-axis elasticity ($c_{33}$) of 241 GPa[25], to estimate $k_t^2$ through:

$$k_t^2 \approx \frac{d_{33,f}^2 c_{33}}{d_{33,f}^2 c_{33} + \varepsilon_0 \varepsilon_{33}} \qquad (1)$$

The quadratic increase in $d_{33,f}$ combined with the nearly constant permittivity, leads to an increase in $k_t^2$ from 13.8% to 76.2% after annealing. The large $k_t^2$ of annealed ScAlN sample is approaching the 85.5% value that is recently reported by Yao et. al.[44] for twisted bilayer Lithium Niobate structure in a 1 mm bulk layer. It is worth noting, the results reported herein are from a 100 nm ScAlN film that is 10,000× thinner, thereby promising extreme frequency scaling of bulk acoustic wave resonators for integrated microwave spectral processing in centimeter and millimeter-wave frequencies. The very large $k_t^2$ of the annealed ScAlN film enables creation of ultra-wide bandpass filters that are excessively needed for beyond-5G wireless communication technologies.

## 2.2. Morphological and structural characterization

To achieve maximum piezoelectric response from a wurtzite crystal, it is important to obtain a highly ordered *c*-axis orientation and columnar grains. ScAlN films deposited by radio-frequency (RF) sputtering often exhibits uneven orientations, resulting in degraded crystalline quality and piezoelectric response.[26] Furthermore, annealing can significantly alter the stress in the film, thereby affecting the crystalline quality and the piezoelectric response of the film.[26,45] To analyze the effect of annealing in boosting crystallinity of the as-grown ScAlN films, we conducted X-ray diffraction (XRD) analysis on ScAlN samples annealed at various temperatures for a duration of 2 hours. The full width at half maximum (FWHM) of the (0002) plane ScAlN ω-peak is 2.85° for the as-grown sample. In **Figure 4(a)**, the trend reveals a consistent decrease in the FWHM as the annealing temperature rises, suggesting a progressive enhancement in crystalline quality. Specifically, the FWHM reaches its minimum value of 2.4° at an annealing temperature of 700 °C, aligning closely with the observed improvement in $d_{33}$. However, as the annealing temperature is raised to 800 °C, there is a sharp reversal in this trend, with the FWHM experiencing a notable increase. This notable rise in FWHM continues to escalate with increasing annealing temperature, rendering it unmeasurable at 900 °C, indicating a substantial loss of crystallinity at such elevated temperatures.

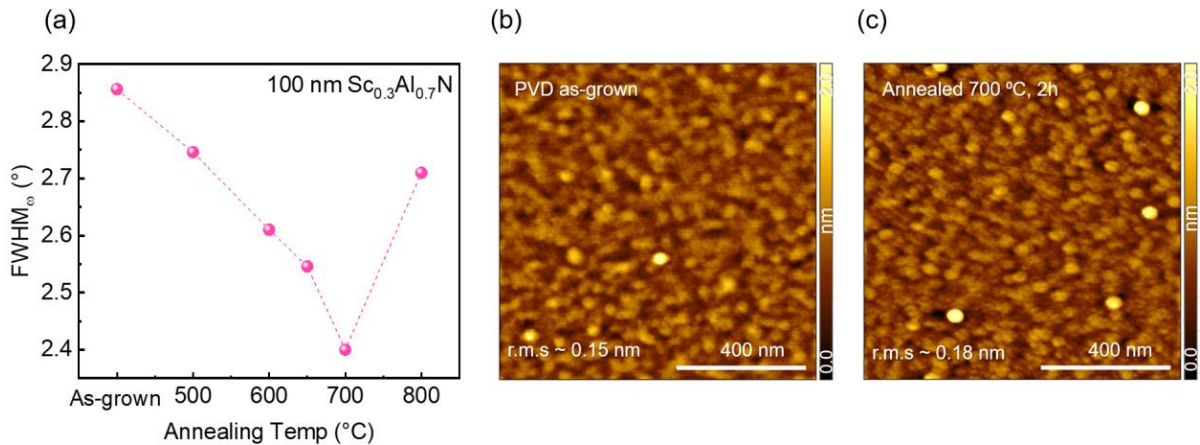

**Figure 4:** (a) Plot of full width at half maximum (FWHM) of the (0002) plane ScAlN ω-peak showing a decreasing trend with increasing annealing temperature, with the lowest value of 2.4° at an annealing temperature of 700 °C, aligning closely with the observation in $d_{33}$ enhancement. Comparison of surface morphology from AFM images obtained for (b) as-grown ScAlN sample showing an r.m.s. roughness of 0.15 nm and (c) sample annealed at 700 °C for 2h, showing an r.m.s roughness of 0.18 nm for a 1 μm × 1 μm scanning area.

Previous studies showed that high temperature annealing was often associated with a degradation of surface morphology, which can translate to degradation in device performance.[46] Thus, to analyze the effect of high temperature annealing on the surface morphology, we compared the atomic force microscopy (AFM) images of pristine and post-annealed samples, at the optimized condition (700 °C, 2 hours), shown in **Figures 4(b)** and **(c).** No significant deterioration could be observed for samples annealed under optimized conditions, with the as-grown ScAlN sample showing an r.m.s. roughness of 0.15 nm and the sample annealed at 700 °C for 2h showing an r.m.s roughness of 0.18 nm for a 1 µm × 1 µm scanning area.

The PFM data presented in **Figure 5** showcases the comparison of ScAlN sputtered film before and after annealing at 700°C for 2 hours. The amplitude plot of as-grown film, which measures the out-of-plane piezo response, exhibits a relatively disordered domain structure with numerous small and randomly oriented unresponsive spots for the as-grown sample. The high density of dark spots suggests significant non-uniformity in piezoelectric activity, indicating that the domains are not well aligned along the *c*-crystallographic direction. Post-annealing, the amplitude image shows a marked reduction in the density of dark spots. Moreover, the overall amplitude response is also higher than the as-grown sample, as can be observed in the scale bar of amplitude plots. This indicates an increased alignment of domains towards the *c*-axis, resulting in the reduction of unresponsive grains and a more homogeneous distribution of piezoelectric activity. The thermal treatment induces beneficial modifications in the material's microstructure by primarily reducing defects and dislocations within the crystalline structure.[47,48] Additionally, annealing mitigates residual stresses introduced during the deposition process, thereby improving mechanical stability and optimizing the stress profile within the film.[49-51]

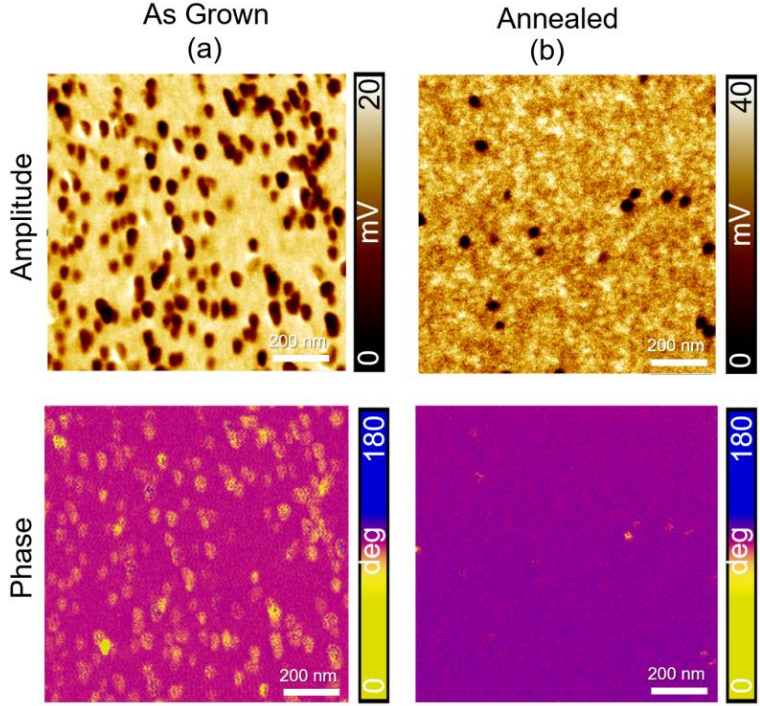

**Figure 5:** Piezoresponse Force Microscopy (PFM) amplitude and phase images of ScAlN sputtered films. (a) As-grown and (b) after annealing at 700°C for 2 hours. The amplitude images show a transition from a disordered domain structure with randomly oriented domains in the as-grown film to a more consistent and aligned arrangement of domains after annealing. The phase images further support this observation, demonstrating a shift from a disordered to a homogeneous arrangement of domains.

The phase images further substantiate these findings by providing insights into the polarization direction and uniformity of domain orientation. The phase image of the as-grown film shows multiple spots with non-uniform phase contrasts, indicating a lack of uniform polarization direction among the domains, contributing to a lower piezoelectric response. Post-annealing, the phase image displays a more uniform phase contrast, with significantly reduced number of mis-oriented domains, indicating coherent orientation along the $c$-axis. Furthermore, annealing stabilizes the desired wurtzite phase of ScAlN, ensuring a homogeneous distribution of Sc within the AlN matrix and reducing the presence of secondary phases that adversely affect piezoelectric performance.[52] Analogously, in more traditional oxide ferroelectrics like hafnium oxide, the material often crystallizes into multiple phases during the growth process that are not desirable for ferroelectric applications. To achieve the transition from centrosymmetric tetragonal phase to a non-centrosymmetric orthorhombic phase, the material is subjected to specific growth conditions or post-deposition treatments such as annealing.[53] The effect of annealing, as explored herein, can be extended beyond nitride ferroelectrics to have a profound impact on the piezoelectric properties of other materials.

In addition to the more homogeneous domain structures observed at a mesoscopic scale, there is often a strong correlation between piezoelectricity of a material and its microscopic structures.[22] Therefore, gaining a deeper insight into piezoelectric properties in wurtzite materials at the atomic scale could potentially establish design guidelines to boost the piezoelectric performance of wurtzite structures. This could allow for the identification of a manageable fundamental parameter that can be further engineered to enhance the wurtzite piezoelectricity, leveraging advanced approaches such as first-principles and machine learning techniques,[22,23] which have garnered significant attention recently. To explore the structural origins of the enhancement in $d_{33}$ after thermal annealing, we performed atomic resolution measurements of the lattice parameters using scanning transmission electron microscopy (STEM). The piezoelectric response along the *c*-axis in wurtzite crystals is defined by the piezoelectric modulus $d_{33}$ that can be approximated as[1,54,55]:

$$d_{33} \approx \varepsilon_0 e_{33} \chi_{33} \qquad (2)$$

Thus, the enhancement in the piezoelectric coefficient $d_{33}$ can be achieved by increasing the polarization-strain coupling ($e_{33}$) and by increasing the polarization-electric field coupling ($\chi_{33}$). Prior studies have shown that the reduction of the lattice parameter ratio (*c/a*) due to Sc substitution in AlN is strongly correlated with an increasing $e_{33}$ and an increase $\chi_{33}$.[55,56] This behavior suggests that other factors, such as compressive in-plane strains or internal stresses that increase the c/a ratio, ratio may be the potential limiting factors of piezoelectric response.[57]

Herein, atomic resolution differentiated differential phase contrast (dDPC)-STEM imaging is used to reliably perform bond length measurements as the images are linearly related to the Laplacian of the phase of the sample transmission function, providing reliable measurement of atomic-column locations of both relatively heavy and light elements.[58,59] Using gaussian fitting of the metal and nitrogen positions, the average values of the *a* parameter, *c* parameter, and *c/a* ratio were measured from atomic resolution dDPC-STEM images of ScAlN before and after annealing at the optimized condition of 700°C for 2 hours, as shown in **Figure 6**. These measurements revealed that the *c/a* ratio is reduced from 1.542 to 1.522 after annealing. This lattice parameter ratio reduction results mainly from an increase in the *a* parameter (3.230 to 3.265 Å) as the decrease in the *c* parameter (4.980 to 4.968 Å) is less significant.

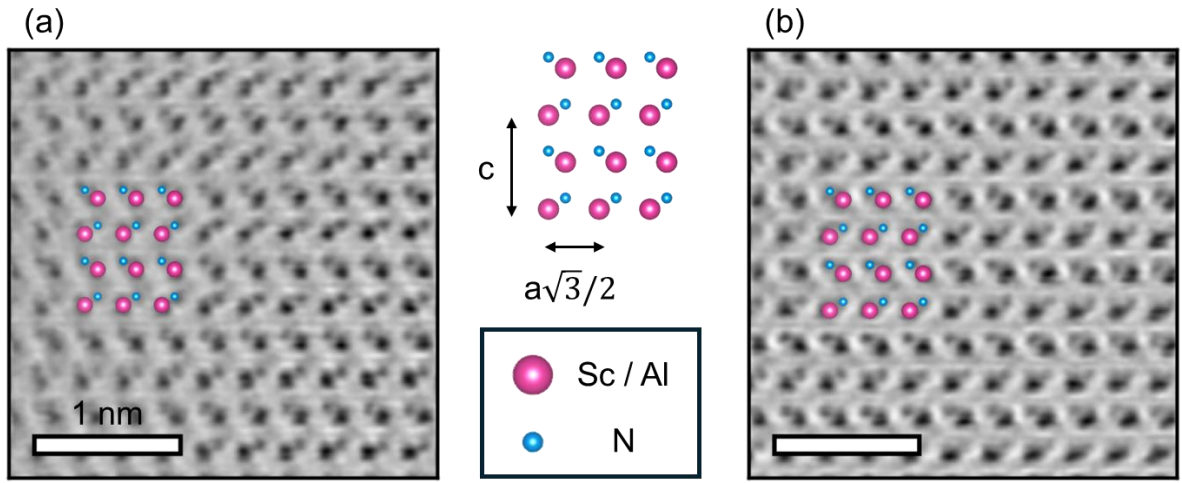

**Figure 6:** dDP*C*-STEM images of the (a) PVD as-grown and (b) the post-annealed sample at the optimized condition of 700 °C, 2 hours.

As the atomic resolution images are localized measurements from the center of the ScAlN thin film, we also performed four-dimensional (4D)-STEM measurements over a larger 550 nm wide area to further confirm the findings of the atomic-resolution dDPC-STEM imaging and provide a more statistically reliable analysis of the lattice parameter changes. **Figure 7(a)** is a virtual dark field image of the ScAlN thin film constructed by the integration of the 4D-STEM data showing columnar grains of 5-30 nm in diameter. Averaging all diffraction patterns from the ScAlN film gives the patterns as shown in **Figure 7(b)**, which displays evidence of a highly ordered *c*-axis growth orientation with a randomly oriented *a*-axis. For reliable measurement of the *a* and *c* parameters, only the diffraction patterns which lie close to the [$11\bar{2}0$] zone axis was used by selecting those with high intensity in the virtual dark field image constructed from integration of the ($1\bar{1}0l$) and ($\bar{1}10l$) diffraction spot positions. The values of *a* and *c* parameters were then measured from each of these electron diffraction patterns using the exit wave power cepstrum (EWPC) transform which provides a precise measurement of lattice parameter values and is more robust to small variations in sample tilt than direct measurement of the diffraction patterns.[60] An example EWPC pattern from the 4D-STEM dataset is given in **Figure 7(c)**, where the spots in the pattern represent the real space lattice of the ScAlN crystal, with periodicities of *a* and *c* in the horizontal and vertical direction, respectively. The distributions of values obtained from the 4D-STEM EWPC analysis of both the as-grown and annealed samples are given in the histograms in **Figure 7(d-f).** The EWPC measurements show that after annealing at the optimized condition of 700°C for 2 hours, the *c/a* ratio is reduced from 1.560 to 1.553 with the *a* parameter increasing from 3.185 to 3.201 Å and the *c* parameter showing little change (4.967 to 4.970 Å). Moreover, the entire histogram for the *a* parameter shows a shift towards higher values, representing an in-plane relaxation of the lattice, that is the driving factor for the *c/a* decrease. The statistical analysis has been summarized for both measurement techniques (refer to Table I in Supplementary Material). Previous studies show using XRD and 4D-STEM measurements that the *a* parameter increases and the *c* parameter decreases with increasing film thickness for $Sc_{0.28}Al_{0.72}N$ grown on GaN, which is attributed to stress relaxation[61]. Thus, our lattice parameter measurements suggest that the enhancement of $d_{33}$ after thermal annealing is likely due, in part, to stress relaxation of the crystal lattice.

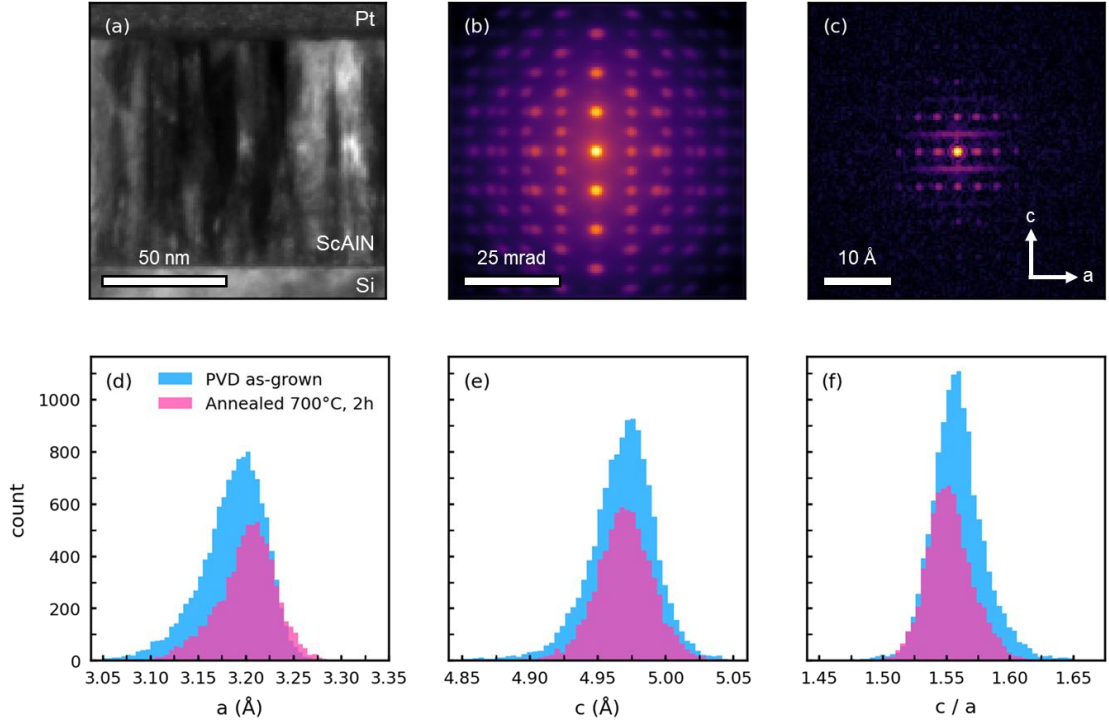

**Figure 7:** (a) Virtual dark field image constructed from the 4D-STEM data by integration of the $(1\bar{1}0l)$ and $(\bar{1}10l)$ diffraction spot positions. (b) Average diffraction pattern from the ScAlN. (c) EWPC of a single 4D-STEM diffraction pattern. Histograms of the lattice parameter values measured via 4D-STEM EWPCs for the (d) *a* lattice parameter, (e) *c* lattice parameter, and (f) lattice parameter ratio (*c/a*).

An important consideration herein is that the values reported in this work represent $d_{33,f}$ and not pristine $d_{33}$, as has been commonly used interchangeably in literature. Although the enhancement observed in $d_{33}$ is related to the in-plane relaxation and an increased *a* lattice parameter of ScAlN upon annealing, we need to consider that it is still limited by the clamping effect from the substrate.

Thus, the true $d_{33}$ value of the material may be much higher when released from the substrate, thus making it even more attractive for applications involving free-standing piezoelectric material systems. The simplicity of this approach can lead to the development of highly scalable and universal processes for enhancement of piezoelectricity across various material systems. Furthermore, this approach opens a promising path to attain even higher $d_{33}$ values by optimizing the deposition process, increasing Sc composition, or the incorporation of other IIIB elements, *e.g.,* Y and La in the wurtzite lattice, transforming piezoelectric and electroacoustic devices over a broad range of applications. In particular, the enhanced $k_t^2$ enabled by the $d_{33}$ boost paves the way for the creation of ultra-wide band filters for the next generations of wireless communication. The large coupling despite the small thickness of 100 nm enables piezoelectric transduction of acoustic waves in low-loss substrates (*e.g.*, silicon, silicon carbide, diamond) with minimum perturbation of quality factor ($Q$). This facilities creation of integrated high-$Q$ resonators and acoustic waveguides for ultra-stable clock generation, high resolution sensors, and emerging optomechanical and quantum-acoustic devices.

## 3. Conclusion

In summary, we have experimentally demonstrated a simple, scalable, and highly effective strategy to dramatically enhance the piezoelectric response of ScAlN by employing a high temperature annealing technique. The atomic level investigation not only advances our fundamental understanding of $d_{33}$ enhancement but also paves the way for the rationale design and optimization of materials with enhanced piezoelectric performance. Moreover, the simplicity of the approach promises scaling accessibility across various material systems. The unprecedentedly high piezoelectric response achieved with ScAlN in this study, along with its potential for seamless compatibility with back-end CMOS integration, not only positions ScAlN but also other IIIB-doped nitride semiconductors as versatile materials for a broad spectrum of applications, spanning from electronics and acoustics to photonics, quantum technologies, and energy harvesting systems, driving progress for commercial adoption.

## 4. Experimental Section

*Synthesis of ScAlN:* A 100 nm $Sc_{0.3}Al_{0.7}N$ was grown by reactive magnetron sputtering on Si (001) substrates with a chuck temperature of 350 °C.

*Annealing:* Initially the samples underwent a thorough solvent cleaning procedure using acetone, isopropyl alcohol (IPA) and de-ionized (DI) water, followed by annealing under various ambient conditions using an *Angstrom Engineering* (AE) Furnace. At first, the sample was placed into the annealing tube, and then the tube was evacuated to eliminate remaining air. The temperature was subsequently increased. After the temperature reached a steady state, the sample was subjected to baking for the intended duration.

*Piezoelectric modulus ($d_{33}$) and capacitance measurements*: Conductive silver paste was used to connect the exposed ScAlN layer to a metal sheet, enhancing signal transmission and measurement precision. PFM investigations utilized vertical domain modes with voltage applied between the top and bottom electrodes. The piezoelectric coefficient $d_{33}$ was determined from the amplitude of the piezoresponse signal using localized surface plasmon resonance (LSPR) amplitude plots. Bias voltage was sequentially applied at 0 V and 10 V, with phase and amplitude responses recorded and background noise subtracted from the 10 V amplitude response. The $d_{33}$ measurements from PFM were further correlated and validated with measurements utilizing a LDV system with a helium-neon laser ($\lambda$=633 nm) from Polytec GmbH, coupled with a Precision Multiferroic Tester from Radiant Technologies, Inc., to electrically stimulate the films. To minimize ambient vibrational noise and ensure a high signal-to-noise ratio (SNR), the setup was placed on an active vibration-isolation platform within an acoustic enclosure. A custom electrode was patterned by standard photolithography on the ScAlN surface with a diameter of 50 µm, and electrical excitation was achieved using tungsten probes with a 10 V amplitude at 20 kHz. The substrate's backside was grounded electrically, and surface displacement was recorded post-bias application. Bias-dependent measurements confirmed consistent displacement values, indicating negligible ambient noise impact on measurements.

Longitudinal displacement butterfly loop was measured using Polytech NLV-2500 Laser Vibrometer, connected to Radiant PiezoMEMS tester providing electric stimulus. Sample for this measurement is n-doped Si/ AlScN/ Mo capacitor with an area of 0.0025 cm$^2$. Testing signals correspond to 60V, bipolar, 2kV triangle waves. To get rid of environmental noise and substrate bending effect, samples were placed on Thorlab vibration isolation table and bonded to large mass. The results were post-processed by averaging five measurement cycles and correcting for drift.

Capacitor-voltage loops are measured by Radiant PiezoMEMS taster, using 500mm × 500mm capacitors, driven by 1MHz triangular pulses with 27 V and 25 V amplitudes for as-grown and annealed samples, respectively.

*Structural and morphological characterization:* For structural characterization of the sample, X-ray rocking curve scans (XRC) were performed using a Rigaku SmartLab diffractometer with a Cu K$_{\alpha 1}$ radiation x-ray source (1.5406 A˚). The surface morphology of the films was analyzed using a Bruker ICON Atomic Force Microscopy (AFM) system. Cross sectional TEM samples of the ScAlN thin films were prepared using a FEI Helios G4 PFIB CXe Dual Beam FIB/SEM, with the final polish using an acceleration voltage of 5 kV. dDPC-STEM imaging was performed using a Themis Z (Thermo Fisher) C$_s$ probe corrected microscope with an acceleration voltage of 200kV and a probe semi convergence angle of 25 mrad. dDPC-STEM imaging was performed using a pixel dwell time of 20 µs and a collection angle of 18 – 69 mrad on a segmented detector with 4 azimuthal segments (DF4). The dDPC images were post-filtered using a gaussian filter with a sigma value of 1.5 pixels. The 4D-STEM data was acquired with an electron microscope pixel array detector (Thermo Fisher) using a pixel dwell time of 1 ms, an acceleration voltage of 200kV, and a probe semi convergence angle of 0.9 mrad, resulting in a probe size with a full-width at half-maximum of approximately 1.5 nm.

**Supporting Information**

Additional information supporting the findings of this work can be found here:

**Acknowledgements**

S.M. and M.M.H.T contributed equally to this work. This work was supported by the Department of Defense Advanced Research Projects Agency (Award # HR00112390018), University of Florida and College of Engineering, University of Michigan. The authors acknowledge the Research Service Centers at the Herbert Wertheim College of Engineering at the University of Florida for the electron microscopy work.

# Supplementary Information

**Table I:** Summary of lattice parameter measurement averages ± one standard deviation.

| Technique | Sample | $a$ (Å) | $c$ (Å) | $c/a$ |
|---|---|---|---|---|
| **dDPC-STEM** | As-grown | 3.230 ± 0.045 | 4.980 ± 0.086 | 1.542 ± 0.033 |
|  | Treated | 3.265 ± 0.052 | 4.968 ± 0.073 | 1.522 ± 0.036 |
| **4D-STEM** | As-grown | 3.185 ± 0.037 | 4.967 ± 0.028 | 1.560 ± 0.022 |
|  | Treated | 3.201 ± 0.031 | 4.970 ± 0.020 | 1.553 ± 0.018 |

The lattice parameter values obtained from both atomic resolution dDPC-STEM imaging and 4D-STEM EWPC analysis are summarized in **Table I**. The two measurement methods are in good agreement with each other and show a clear trend of a reduction in the $c/a$ ratio, primarily driven by an increase in the $a$ parameter while the c parameter is less significantly changed. Additionally, the 4D-STEM measurements show a reduction in the distribution of measured lattice parameter values after thermal annealing, consistent with the reduction in FWHM of the (0002) plane ScAlN ω-peak measured with XRD indicating an improvement in the crystalline quality of the sample.